UDC 621.391.1

# The function space to describe the dynamics of linear systems

V.N. Tibabishev

Asvt51@narod.ru

Usually, the dynamics of linear time-invariant systems described by an integral operator of convolution type, which is defined in the Hilbert space of Lebesgue square integrable functions on the whole line. Such a description leads to contradictions.  It is shown that the transition to the Hilbert space of almost periodic functions leads to the elimination of the detected inconsistencies.  Multiple signals and interference with discrete spectrum are systems of sets. The properties of these systems lead to a new more effective method to combat noise in this space. The method used to identify the differential equations for the airbus. Baseline data were obtained during automatic landing.

## 1. Introduction

Initially restrict ourselves to the simplest case, when a one-dimensional dynamical system is linear and stationary. Authors of many studies suggest that the observed input and output of the system such signals are stationary ergodic random processes [1, § 7.4]. In this cross-correlation function $R_{yx}(\tau)$, the autocorrelation function $R_{xx}(\tau)$ the input signal $x(t)$, output signal $y(t)$ and the weight function $k(t)$ are interconnected by an integral operator of convolution type defined in the Hilbert space of Lebesgue $L_2(-\infty, +\infty)$ [1, § 9.6].

$$A_x k \equiv \int_{-\infty}^{+\infty} R_{xx}(t-\tau)k(\tau)d\tau = R_{yx}(t). \quad (1)$$

The kernel is generated by the autocorrelation function, which is diagonally at $t = \tau$ an infinite plane $-\infty < t, \tau < +\infty$ nowhere decreases from its maximum positive value. Therefore, this operator is not bounded Hilbert-Schmidt operator, since the double integral $\iint_{-\infty}^{+\infty} R_{xx}^2(t\text{-}\tau)dtd\tau$ diverges. The author got this information from Professor MGTU Bauman V.F. Biryukov now deceased.

For unbounded operators, the right-hand side of equation (1) may not belong to the Hilbert space of Lebesgue $L_2(-\infty, +\infty)$. This eliminates the correct application defined in a Hilbert space $L_2(-\infty, +\infty)$ Fourier transforms for solving integral equations in the frequency domain.

Averaging over many realizations cannot be performed in most cases. It is believed [1, § 7.4], that a stationary random process has an ergodic property. In this case, for example, the correlation function of a random stationary process is found by time-averaging a single realization

$$R_{xx}(\tau) = \lim_{T \to \infty} \frac{1}{2T} \int_{-T}^{+T} x(t-\tau)x(t)dt. \qquad (2)$$

It is known [2, § 7.11], a random stationary process has an ergodic property, when, and only if its spectral function is continuous. Such random processes are elements of the Hilbert space of Lebesgue $L_2(-\infty, +\infty)$ [2, § 7.5] and satisfy the condition

$$\int_{-\infty}^{+\infty} x^2(t)dt < \infty. \qquad (3)$$

Using (3) in (2), we find that well-known algorithm for solving applied problems for random stationary ergodic processes leads to the trivial case when the operator equation (1) $R_{xx}(\tau) = 0$ and $R_{yx}(\tau) = 0$ on the whole line.

Apparently, such a paradoxical situation arises for the reason that in dealing with the problem of representing signals in a linear time-invariant systems have been used incorrect assumptions.

It is known [3, § 3] that the random component of the stationary process with discrete spectrum belongs to the Hilbert space of almost periodic functions. Contradictions do not appear above, if we use the Hilbert space of almost periodic functions. In this regard, we consider the properties of systems of sets of signals in the Hilbert space of almost periodic functions with the following conditions.

First, instead of exact initial data input $x(t) \in X$ and output $y(t) \in Y$ are given approximate initial data in the form of systems of sets of input signals $\mathfrak{A}_{xn} = X \cup N$ and a system of sets of output signals $\mathfrak{A}_{ym} = Y \cup M$, distorted by an additive noise $n(t)$ and $m(t)$ as $\tilde{x}(t) = x(t) + n(t)$ and $\tilde{y}(t) = y(t) + m(t)$.

Secondly, the integral operator of convolution type differential operator answers the $n$ −th order

$$Dy \equiv \sum_{k=0}^{k=n} a_k y^{(k)}(t) = x(t), \qquad (4)$$

where $y^{(k)}(t)$ - a derivative of $k$ − th order.

The operator of differentiation is the inverse with respect to the integral operator $D = A_x^{-1}$.

Third, the direct and inverse operators are defined only for the exact original data. It follows from the fact that the noise $n(t)$ does not generate involuntary movements in the control system and the noise $m(t)$ is not part of the forced motion.

## 2. The space of functions for linear time-invariant operators

Typically, the control system are multi-dimensional non-linear and time-dependent, such as aircraft. In the motion of an airplane, for example, the final approach speed and its weight varies little across the range of planting. For other flight regimes can be distinguished intervals at which the plane can be approximately assumed to be linear and stationary for a given mode of flight.

In many cases, control systems are systems with correlated inputs. The article shows how to bring this system to a multi-dimensional system with uncorrelated effects. This technique greatly simplifies the tasks of multidimensional systems can be considered as a multidimensional system as a union of independent linear one-dimensional steady-state control channels. We shall therefore consider the one-dimensional linear time-invariant control system

As the dimensions of the aircraft, carrying capacity and speed of flight airframe subject to elastic deformations, which strongly influence the management process. Therefore, the operators of differentiation (4) contain derivatives of higher orders.

In describing the trajectory of the aircraft in space in the earth coordinate system, consider a material point of the plane. The influence of the size of aircraft and airframe elastic deformation is not taken into account. The aircraft has six degrees of freedom. The translational motion of the aircraft, for example, one space component is described by second order differential equation of Newton

$$m\ddot{y}(t) = x(t), \qquad (5)$$

where $\ddot{y}(t)$ - the second derivative of the output signal on the selected channel control, $x(t)$ - the force interaction at the entrance.

From this equation it is easy to obtain the operator equations for the eigenfunctions and eigenvalues $m\ddot{y}(t) = ay(t)$. It follows that the eigenfunctions for the operator equation are functions $y_i(t) = \exp(j\omega_i t)$.

Associated coordinate system is used to describe the motion of parts of the aircraft relative to the center of mass. In addition, as a rule, the operator of differentiation is of high order. It is not hard to make sure that the eigenfunctions for equations (5) and (4) does not depend on the order of the differential equation.

The angular frequency ω is related to the cyclic frequency $f$ by the equation $\omega = 2\pi f$. If the cyclic frequency belongs to the set of real numbers, the angular frequency by the factor π belongs to a subset of transcendental irrational numbers. A subset of the angular frequency does not contain a subset of rational numbers. A subset of transcendental numbers has an infinite countable set of discontinuities. Such concepts of mathematical analysis, the derivative, differential and integral do not exist on non-continuous independent variable ω.

The well-known integral representation of the $i$-th realization of a random process, for example, $x_i(t)$ can be formally represented in the form of a continuous independent variable cyclic frequency [1, § 4.3]

$$x_i(t) = \int_{-\infty}^{+\infty} C_{xi}(2\pi f) \exp(j2\pi f t)\, df. \qquad (6)$$

Since the function $C_{xi}(2\pi f)\exp(j2\pi f t)$ have jumps in the interval $-\infty < f < \infty$ is infinite, but countable by the power of rational numbers jumps, instead of the Riemann integral should be taken Lebesgue-Stieltjes integral in infinite measure. It is known [4, § 4.4], that this Lebesgue-Stieltjes integral is a Fourier series, defined by $f_k$ of the integrand

$$x_i(t) = \sum_{k=-\infty}^{k=+\infty} C_{xi}(2\pi f_k) \exp(j2\pi f_k t), \qquad (7)$$

where the values of $C_{xi}(2\pi f_k)$ are defined by a Stieltjes integral of a function $g(t) = t/2T$

$$C_{xi}(2\pi f_k) = \int_{-\infty}^{+\infty} x_i(t) \exp(-j2\pi f_k t)\, dg(t) = M\{x(t)\exp(-j2\pi f_k t)\, dg(t)\} \equiv$$

$$\equiv \lim_{T \to \infty} \frac{1}{2T} \int_{-T}^{+T} x_i(t) \exp(-j2\pi f_k t)\, dt. \qquad (8)$$

It follows that a stationary random process can not have a continuous spectrum.

For convergent series (7) $x_i(t)$ is an element of a separable subset of Besicovitch almost periodic functions $B_2(-\infty, +\infty)$ with a discrete spectrum [5]. In addition, each bounded solution of equation (4) belongs to the set of almost periodic functions $y_i(t) \in B_2(-\infty, +\infty)$ [6, § 9.5].

It is known [7, § 5.62], that on the set of almost periodic functions defined by integral completely continuous normal operator of convolution type, for example,

$$y_i(t) = M\{x_i(t-\tau)k(\tau)dg(t)\} = \sum_{d=-\infty}^{d=+\infty} C_{xi}(2\pi f_d) s_k(2\pi f_d) \exp(j2\pi f_d t), \quad (9)$$

displaying $B_2(-\infty, +\infty)$ into itself.

A simple test of time-averaging, for example, to model the input signal $\hat{x}_i(t) = C + x_i(t)$, we find that a stationary random process with discrete spectrum and a non-zero expectation of $C$ has an ergodic properties of the first order of expectation and centered stochastic process (7) does not have the ergodic property of second-order dispersion and correlation functions. The operator of convolution type (2) is defined in the Hilbert space of almost periodic functions, but its value depends on the number of implementation, for example, the input $x_i(t)$. Therefore, the operator (2) does not define a deterministic correlation function.

### 3. Frequency analysis of signals with discrete spectrum

The well-known correlation method of dealing with noise is unacceptable for stationary random processes with discrete spectrum, for two reasons. First, such random processes do not have the ergodicity of the second order. Second, even with a very rare opportunity of averaging over many realizations of the autocorrelation function does not suppress the additive uncorrelated noise, distorting the input signal.

Therefore, to control noise we propose a new method of frequency control noise instead of inaccessible and poor correlation method. Frequency method of analysis differs from the known spectral analysis that the frequency analysis is preceded by spectral analysis. If, for example, the amplitude spectral analysis is to determine the dependence of the amplitudes of the harmonic components as a function of frequency, the frequency analysis is to determine only the set of harmonic frequencies of the selected sets.

In the Hilbert space of Lebesgue $L_2(-\infty, +\infty)$ frequency analysis problem does not arise. In this space, the spectrum is continuous. Therefore, any random frequency taken $\omega_k \in \mathfrak{M}$ is contained in each set of frequencies of the system sets the frequency of the harmonic components $\mathfrak{M} = \Omega_x \cup \Omega_n \cup \Omega_y \cup \Omega_m$, where the sets of $\Omega_x$-frequency harmonic components of the exact input signals $x(t)$, $\Omega_n$- set the frequency of the harmonic components noise, distorting the input signal $n(t)$, $\Omega_y$ - set the frequency of the harmonic components of accurate output signals $y(t)$ and $\Omega_m$- set frequency harmonic noise, distorting the output signal $m(t)$. In the space of periodic signals of the first harmonic frequency is determined by the repetition period for the entire set of frequencies. The frequencies of the higher frequencies are multiples of the frequency of the first harmonic frequency for each subset of the sets of frequencies .

In the space of almost periodic functions of the spectra of the signals are discrete and disparate frequency of harmonics. The frequency of each harmonic component should be determined separately. As shown below, the frequency of the harmonic components of precise signals do not coincide with the frequencies of harmonic components of the additive noise.

Consider one of the ways to determine the frequency of the harmonic components. Let realization, for example, the input signal distorted by additive noise. Initially, local maxima in the amplitude frequency response estimates are the frequencies for multiple set of frequencies. The frequency of the first harmonic is selected by the duration of implementation. From the resulting set of emitted frequencies harmonic components, whose energy is less than the specified level. The method of successive approximations in the vicinity of the maxima obtained at multiple frequencies is updated frequency values for the global maximum. In this way, the system sets are approximate estimates of the frequency $\mathfrak{W}_{\tilde{x}} = \Omega_x \cup \Omega_n$. In a similar way there is an approximate evaluation of the system sets the output signal $\mathfrak{W}_{\tilde{y}} = \Omega_y \cup \Omega_m$. The resolution of the frequency analysis is $\Delta\omega = 2\pi/T$, where $T$ -duration of implementation.

We define the properties of systems of sets of signals and systems sets the frequency of the harmonic components of stochastic processes with discrete spectrum belonging to the Hilbert space of almost periodic functions $B_2(-\infty, +\infty)$.

Power orthonormal system of functions $\exp(j\omega t) \in \mathbb{M}$ is a continuum and determined by the capacity of the set of transcendental irrational numbers of the angular frequency. It is known [7, § 1.10], that what would have been the power of an orthonormal system $\mathbb{M}$ in an arbitrary Hilbert space, every vector x has at most a countable set of nonzero projections on the elements of . It follows that in any Hilbert space of vector $x_i$ is not a Fourier integral (6), and Fourier series (7).

When solving practical problems it is necessary to consider systems of sets. Let the system of sets of input signals $\mathfrak{A}_{xn} = X \cup N$, and the system sets the output signals $\mathfrak{A}_{ym} = Y \cup M$. Purely formal and can be other systems of sets, for example, $\mathfrak{A}_{xx} = X \cup X$, $\mathfrak{A}_{xym} = X \cup Y \cup M$, $\mathfrak{A}_{xnm} = X \cup N \cup M$.

In the theory of stochastic processes distinguish correlated and uncorrelated processes. Clearly, the system sets $\mathfrak{A}_{xx}, \mathfrak{A}_{nn}, \mathfrak{A}_{yy}$ and $\mathfrak{A}_{mm}$ are correlated systems of sets. Theoretically, by averaging over many realizations can find the autocorrelation function of a random stationary process $x_i$

$$R_{xx}(\tau) = \sum_{k=-\infty}^{k=+\infty} \sigma^2 C_i(\omega_k) \exp(j\omega_k \tau), \qquad (9)$$

where $\sigma^2 C_i(\omega_k)$ - the variance of the random amplitudes of harmonic components with frequency $\omega_k$.

From a comparison of expressions (6) and (9) we find a first property. If multiple implementations $x_i(t) \in X$ gives rise to the correlation function of the form (9), each stochastic process is a linear hull of deterministic orthonormal basis of the correlation function $exp\ (j\omega_k\tau)$ with random coefficients. In other words. Each process $x_i(t) \in X$, there corresponds a definite set of harmonic frequencies $\omega_k \in \Omega_x$, no matter the number of implementation.

All antiblackout by Wiener based on the assumption that the additive noise, distorting the accurate output signals are uncorrelated with the exact input signals, it is assumed that the elements of sets, for example, $\mathfrak{A}_{xm} = X \cup M$, cross-correlation function is zero$R_{xm} = \overline{x_l m_l} = 0$, where the bar denotes averaging over an infinite set of realizations.

The main disadvantage of the correlation method is that it is impossible to get a set of signals under the same conditions. For example, you cannot do a lot of automatic landing aircraft on a runway under the same weather conditions. This eliminates the possibility of a correct determination of the correlation functions on the set of realizations.

Therefore, instead of using the correlation functions we have to use other criteria for estimating the relationship between random processes. For example, instead of determining the correlation functions can use the second property of the theoretical linear dependence or independence of random processes. Given a realization of the $i$ -th output signal, distorted by additive noise. Such a realization can be written as

$$y_{mi}(t) = y_i(t) + m_i(t) = \sum_k C_{ki} \exp(j\omega_k t) + D_{ki} \exp(jv_k t).$$

If there are nonzero coefficients $C_{ki}$ and $D_{ki}$, in which $y_{mi}(t) = 0$, $y_i(t)$ and $m_i(t)$ are correlated random processes. Search coefficients satisfy or not satisfy the above condition is not an easy task. Obtain accurate values of the frequencies $\omega_k$ and $\nu_k$ impossible, since their exact values are determined by non-periodic infinite fractions.

It is known [4, § 3.4], that if the components of the process, for example, $y_i(t)$ and $m_i(t)$ are linearly independent processes, they are orthogonal. Implementation of this third theoretical properties checked by the vanishing of the scalar product

$$(y_i, m_i) = M\{y_i(t)m_i(t)dg(t)\} \equiv \lim_{T\to\infty} 1/2T \int_{-T}^{+T} y_i(t)m_i(t)\, dt = 0 \ .$$

The difficulty in applying this most simple criterion is that the exact implementation of the output signal and additive noise $y_i$ are not available for measurement. In reality, initial data are available only to a system of sets $\mathfrak{A}_{ym} = Y \cup M$ and $\mathfrak{A}_{xn} = X \cup N$.

From the first theoretical properties that having a single implementation, for example, the input signal distorted by additive noise can be found a lot of system frequency harmonic functions of the observed process.

Spoiler $n_i(t)$ does not generate a forced movement of the control channel. The exact component $x_i(t)$ and nois $n_i(t)$ have a different nature. Therefore, they are considered independent processes, and hence orthogonal to the third property. Using the representation for the realization of the random process of the form (7), we obtain an expression for the scalar product

$$(x_i, \bar{n}_i) = M\{x_i, \bar{n}_i\} = M\{\sum_{k=-\infty}^{k=+\infty} \sum_{p=-\infty}^{p=+\infty} C_i(j\omega_k) D_i(-j\nu_p) \exp(j(\omega_k - \nu_p)t) dg(t)\}$$

This scalar product is zero if for $\forall k$ and $\forall p$ holds $\omega_k \neq \nu_p$. It follows that if the random processes with discrete spectrum are uncorrelated (linearly independent or orthogonal), the suppression of the frequency sets of orthonormal sets of these processes is empty $\mathfrak{M}_x \cap \mathfrak{M}_n = \Omega_{xn} = \emptyset$, where $\omega_k \in \mathfrak{M}_x$, $\nu_p \in \mathfrak{M}_n$. This is the latest theoretical fourth property, which will most often be used to solve applied problems.

### 3. Examples of applied problems

Meeting the challenges of multidimensional control systems is greatly simplified if the input signals are uncorrelated. We show that the multi-dimensional control system with correlated input effects can lead to a conventional system with independent inputs. Consider a simple example. Let the control system contains two linearly independent log $\mathfrak{A}_{\tilde{x}1} = X_1 \cup N_1$ and $\mathfrak{A}_{\tilde{x}2} = X_2 \cup N_2$. On the first theoretical property that each system meets the system of sets of signals sets the frequency of the harmonic components $\mathfrak{M}_{\tilde{x}1} = \Omega_{x1} \cup \Omega_{n1}$ and $\mathfrak{M}_{\tilde{x}2} = \Omega_{x2} \cup \Omega_{n2}$. Since the sets of signals are uncorrelated (linearly independent), then the pair intersection of all sets of frequencies is empty $\Omega_{x1} \cup \Omega_{n1} \cap \Omega_{x2} \cup \Omega_{n2} = \Omega_{12} = \emptyset$.

If the repetition of experiments at a certain time interval to perform additional coordinated turn aircraft, then at each input will function between the input signals $v(t) \in V$. In this case, a system with correlated input to the $\mathfrak{A}_{\tilde{x}1v} = X_1 \cup N_1 \cup V$ and $\mathfrak{A}_{\tilde{x}2v} = X_2 \cup N_2 \cup V$. Change and the system sets the frequency of the harmonic components $\mathfrak{M}_{\tilde{x}1v} = \Omega_{x1} \cup \Omega_{n1} \cup \Omega_v$ and $\mathfrak{M}_{\tilde{x}2v} = \Omega_{x2} \cup \Omega_{n2} \cup \Omega_v$. At the same time will change and result definitions of the sets of frequencies intersection $\Omega_{x1} \cup \Omega_{n1} \cup \Omega_v \cap \Omega_{x2} \cup \Omega_{n2} \cup \Omega_v = \Omega_v \neq \emptyset$.

We find the intersection of the sets of systems at finite frequency resolution of $\mathfrak{M}_{\tilde{x}1v} \cap \mathfrak{M}_{\tilde{x}2v} = \Omega_v = \{\omega_v \in \Omega_v: |\omega_{x1} - \omega_{x2}| < 2\Delta, \forall \omega_{x1} \in \mathfrak{M}_{\tilde{x}1v}, \forall \omega_{x2} \in \mathfrak{M}_{\tilde{x}2v}, \omega_{x1} \neq \omega_{x2}\}$. This operation is easily implemented by computer software.

The uncorrelated system is constructed using the subtraction operation, if you know a subset of the frequency of communication. We shall call the conditional system $\mathfrak{M}_{\tilde{x}1}$ and $\mathfrak{M}_{\tilde{x}2}$. $\mathfrak{M}_{\tilde{x}1} = \mathfrak{M}_{\tilde{x}1v} \setminus \Omega_v$ and $\mathfrak{M}_{\tilde{x}1} = \mathfrak{M}_{\tilde{x}1v} \setminus \Omega_v$ and $\mathfrak{M}_{\tilde{x}2} = \mathfrak{M}_{\tilde{x}2v} \setminus \Omega_v$, where ·· - a symbol of the subtraction operation. It must be emphasized that conventional ways of uncorrelated signals obtained by calculation and not a substitute for acting on the input source system of correlated input signals.

Consider another example of the formation of correlated signals. In the Hilbert space of almost periodic functions $B_2(-\infty, +\infty)$ defined by a compact normal operator of convolution type

$$y_i(t) = \lim_{T \to \infty} 1/2T \int_{-T}^{+T} x_i(t - \tau) k(\tau) d\tau =$$

$$= \sum_{d=-\infty}^{d=+\infty} K(j\omega_k) C_i(j\omega_k) \exp(j\omega_k t), \qquad (10)$$

where $K(j\omega_k)$ and $C_i(j\omega_k)$ - Fourier exponents defined in the Hilbert space $B_2(-\infty, +\infty)$ by (8).

Comparing this expression with (7) that the implementation of the $y_i(t)$ is a linear combination of the components of the input signal $C_i(j\omega_k)$. At this intersection of the sets of frequencies accurately input and output signals of the exact same set of frequencies of the harmonic components of the nucleus.

The most important task is the task of selection (filtering) the precise input and output signals from interference. We show that the Hilbert space of almost periodic functions, we have *Theorem filtering of signals from noise. Let multivariate linear time-invariant control system have correlated inputs. All or part of the inputs contain independent coupling functions between the inputs. Each control channel is a stationary linear normal completely continuous operator of convolution type defined in the Hilbert space of almost periodic functions, denoted by a serial number input and output. Each component of the exact input, along with the relationship between the input signals are distorted additive uncorrelated noise. Every exact output is generated by one or more accurate input signals and uncorrelated noise corrupted. If you received a conditional system for the front of uncorrelated signals corrupted by uncorrelated noise, the suppression of the union of sets of frequencies of the harmonic components of the system of conditional input signals from a union of sets of frequencies of the harmonic components of accurate output signals with multiple frequency harmonic components of the additive noise output of the selected channel management is a subset of frequencies only accurate harmonic components of an orthonormal system of functions, generating the core part of the operator equation exact linearly independent component of the input signal.*

*The proof.* In multivariate control systems, for example, distinguish between the main plane and cross-channel management. The aircraft is possible to identify a number of inputs: elevator, rudder, ailerons, thrust lever (throttle) and a number of other inputs. In a number of outputs are measured: the pitch, heading, roll, airspeed and other output processes. The elevator pitch is designed to control the aircraft. RED designed to control the air speed, etc. These control channels are straight.

In many cases, the thrust vector offset from the center of mass of the aircraft. At the same time if you do not work on the elevator, and change only the traction motors, you may receive up or down moment nose that will change the pitch of the aircraft. This channel controls the pitch is a cross.

In general, for each selected output q affect the input signals of the direct and cross channels. At the output of q instead of the exact total output signal is observed distorted additive noise

$$\tilde{y}_q(t) = \sum_{l=1}^{l=d} y_{lq}(t) + m_q(t),$$

where $y_{lq}(t)$- an accurate output signal $q$, generated by the input signal at the input of $l$ , and

$$y_{lq}(t) = M\left\{\left(x_l(t-\tau) + \sum_{c=1}^{c=d} v_{lc}(t-\tau)\right)k_{lq}(\tau)dg(\tau)\right\},$$

where $x_l(t-\tau)$- uncorrelated component of the input signal at $l$,

$v_{lc}(t-\tau)$- a communication function between the inputs $l$ and c, for all $l \neq c$;

$k_{lq}(\tau)$ - the weight function of the control channel between the input $l$ and output $q$,

$d$ -number of inputs.

By the first property and the Fourier transform (8), we obtain the system sets the frequency of the harmonic components of output signals and interference

$\mathfrak{M}_{\tilde{y}q} = \cup_l (\Omega_{xl} \cup_c \Omega_{vlc}) \cup \Omega_{mq}$,

where $\Omega_{xl}$ - respectively the set of harmonic frequencies of the exact component $x_l(t)$;

$\Omega_{mq}$ - respectively the set of frequencies of harmonic components of interference $m_q(t)$, the distorting component $y_q(t)$.

$\cup_c \Omega_{vlc}$-, respectively, the union of all the harmonic components of all communication functions $v_{lc}(t)$ between the inputs c and $l$, where c $\neq l, l \in [1, d]$.

Arbitrarily chosen input $p$. We find the conditional system sets the frequency of the harmonic components of independent input signals $\mathfrak{M}_{\tilde{x}p} = \Omega_{xp} \cup \Omega_{np}$. As shown in the example above, the system sets the frequency $\mathfrak{M}_{\tilde{x}p}$ does not contain a set of uncorrelated frequency harmonic components as a liaison $\cup_l (\cup_c \Omega_{vlc})$.

We find the intersection of the sets of conditional frequency of the harmonic components between the input and output $p\ q$. The intersection of all subsets of sets of frequencies of these systems with the exception of one system $\Omega_{xl}$ when $l = p$ is empty. In this case $\mathfrak{M}_{\tilde{x}p} \cap \mathfrak{M}_{\tilde{y}q} = \Omega_{xp}$. This set of frequencies of harmonic components determine the exact basis of the kernel $x_p(t-\tau)$. QED.

Having set the exact frequencies of the input signal can be solved, for example, the task of identifying the selected channel frequency response between the input and output $p\ q$. Fourier exponents $C_x(\omega_k)$ and $C_y(\omega_k)$ the exact components of the input and output signals are only on the set of frequencies $\omega_k \in \Omega_{xp}$ to approximate data

$C_{xp}(\omega_k) = M\{\left(x_p(t) + \sum_{c=1}^{c=d} v_{pc}(t) + n_p(t)\right) \exp(j\omega_k t) \, dg(t)\}$ and $C_y(\omega_k) = M\{\left(\tilde{y}_q(t) + m_q(t)\right) \exp(j\omega_k t) \, dg(t)\}$. Only the exact components of the signals $x_p(t)$ and $y_q(t)$ contain harmonics with frequency $\omega_k \in \Omega_{xpq}$. All other components vary in frequency and so are orthogonal components. Value of the desired frequency response $W_{pq}(j\omega_k) = \frac{C_y(\omega_k)}{C_{xp}(\omega_k)}$.

According to data obtained in automatic landing jumbo jet class of IL-96-300 were obtained, and orders of the coefficients of linear ordinary differential equations of the two control channels: elevator - the pitch and throttle - the pitch of 9 and 5 orders of magnitude. The orders of the differential equation are high due to the influence of elastic deformations.

## Conclusion

The well-known theory of stochastic processes of linear stationary systems based on the assumption that the random stationary processes and their correlation functions are elements of the Hilbert space of Lebesgue on the whole line with a continuous spectrum. Random process with a continuous spectrum has an ergodic property, as stated in the well-known works. Therefore, when solving applied problems of correlation functions are found by averaging the realizations of the time. It is shown that under these assumptions of stochastic processes and correlation functions, there are contradictions.

Khinchin A.Y. showed that random stationary process is an element of the Hilbert space of Bohr with a discrete spectrum. It is shown that the properties of random processes must be considered on a system of exact sets of processes and noise. The correlation function is of the deterministic characteristic stochastic process. It is established that the frequency of the harmonic components are determined characteristic of the random process. If the random processes belong to the same set with a given correlation function.

As a rule, higher-dimensional systems are systems with correlated inputs signals. The method was developed to align systems with correlated inputs to a system with uncorrelated inputs. This method greatly simplifies the solution of the problem in multidimensional systems.

A new method of dealing with noise received on a set of random processes with discrete spectrum. We prove a theorem filtering of signals from noise on the input and output on this set of signals. A new method of dealing with interference leads to a new methods for solving problems of control, such as identification of dynamic characteristics of control systems.

The method used to solve the problem of identification of differential equations of the two control channels according to the passive experiment in automatic landing jumbo jet.

The author is grateful to Associate Professor O.M. Pokolenko for your attention to the article.